%% file: article_final.tex
\RequirePackage{ifpdf}
\documentclass[letterpaper]{article}
\pdfoutput=1

\usepackage[usenames,dvipsnames]{pstricks}
\usepackage{jheppub}
\usepackage[utf8]{inputenc}
\usepackage[english]{babel}
\usepackage[babel]{csquotes}
\usepackage{amsmath}
\usepackage{amsfonts}
\usepackage{amssymb}
\usepackage{mathtools,colonequals}
\usepackage{mathrsfs}
\usepackage{bm}
\usepackage{xcolor}
\usepackage{bbold}
\usepackage{braket}
\usepackage{cases}
\usepackage{graphicx}
\usepackage{graphics,multicol}
\usepackage{tikz}
\usepackage{physics}
\usepackage{soul}
\usetikzlibrary{decorations.pathmorphing}
\usetikzlibrary{decorations.markings}
\usepgflibrary{shapes.geometric}
\usepgflibrary{shapes.symbols}
\usetikzlibrary{arrows.meta}
\usetikzlibrary{shapes.misc}
\usetikzlibrary{fadings}

\tikzset{cross/.style={cross out, draw=black, minimum size=2*(#1-\pgflinewidth), inner sep=0pt, outer sep=0pt},
cross/.default={1pt}}

\newcommand{\vphi}{\varphi}
\newcommand{\<}{\langle}
\renewcommand{\>}{\rangle}

\usetikzlibrary{arrows}

\usepackage{verbatim}
\usepackage{booktabs}
\usepackage{array}

\usepackage{color}
\usepackage{subcaption}
\usepackage[percent]{overpic}
\usepackage{multirow}

\input{defs}

\title{Defect mediated quantum melting of charge ordered insulators}
\author[a, 1]{Abijith Krishnan,}
\author[a, b, 1]{Ajesh Kumar, \note{These authors contributed equally to this work.}}
\author[a]{and T.\ Senthil}

\affiliation[a]{Department of Physics, Massachusetts Institute of Technology, Cambridge, MA  02139, USA}
\affiliation[b]{Institute for Theoretical Physics, University of Cologne, 50937 Cologne, Germany}
\emailAdd{abijithk@mit.edu}
\emailAdd{ajesh.kumar@uni-koeln.de}
\emailAdd{senthil@mit.edu}

\abstract{
Two-dimensional (2d) electronic systems on a lattice at fractional filling $\nu = p/q$ exhibit a competition between charge ordered insulators, called Wigner-Mott insulators (WMIs), at large Coulomb repulsion and Fermi-liquid metals at large electronic kinetic energy. When those two energy scales are roughly equal, insulating states that restore the lattice translation symmetry, which we call quantum charge liquids (QCLs), may emerge. When gapped, these QCLs must exhibit topological order. In this work, we show that the allowed topological ordered phases that are proximate to the WMI strongly depend on the charge ordering in the WMI. In particular, we show that when $q$ is even, no direct transition exists between a WMI with the smallest allowed unit cell size from filling constraints, i.e., the  ``minimal" WMI, and the topological order with the smallest ground state degeneracy on a torus allowed by filling constraints, i.e., the  ``minimal" TO.
Furthermore, we describe the quantum melting transition of the WMIs to the proximate QCLs in terms of the proliferation of the topological defects of the WMIs. 
The field theory of this transition in terms of the topological defects reveals their role as precursors to the anyon excitations in the QCLs.
}

\begin{document}
\maketitle
\newpage
\section{Introduction}
The development of tunable two-dimensional (2d) electron systems has motivated deep theoretical and experimental exploration of strongly-correlated electronic phases of matter. These phases are controlled largely in part by the interplay between Coulomb repulsion and electronic kinetic energy. If the electronic kinetic energy is sufficiently large, electrons on a crystalline lattice form metallic Fermi liquid states. When the energy of Coulomb repulsion dominates and the electronic filling (i.e., average number of electrons per unit cell) $\nu$ is rational, the electrons form an insulating state called a fermionic Wigner-Mott insulator (WMI)\footnote{This state is also called a generalized Wigner crystal in the literature.} and localize into a spatial pattern that spontaneously breaks the crystalline lattice symmetry. Evidence of WMIs has been observed specifically in moir\'e transition metal dichalcogenides (TMDs) \cite{Regan2020_moire, Wang2020_tWSe2, Xu2020_moire_insulators,li2021imaging,huang2021correlated, Mak2022_semiconductor_moire,li2021charge,tang2022dielectric}.

The yet undetermined nature of the transition between 2d WMIs and the metallic Fermi liquid has attracted recent theoretical \cite{musser_et_al_2022} and numerical \cite{Matty2022_moire_Wigner, Macdonald_numerics_2023, zhou_et_al_2024} attention. These references have proposed models and mechanisms for direct WMI to metal transitions (both first order and continuous) and transitions with intermediate states. Moreover, two recent experiments have studied the mechanism through which an insulating electronic crystal may disorder through quantum fluctuations, i.e., melt. 
In MoSe$_2$/WS$_2$~\cite{tang2022dielectric} and biMoSe$_2$ \cite{Zong2025_moSe2_quantum_melting} moir\'e heterostructures, the quantum melting of a WMI was studied using optical and transport measurements respectively, with evidence found for continuous metal–insulator phase transitions.

A related question is: what intermediate states are allowed between the WMI and metal? Conventional weak-coupling theories suggest intermediate states that break the lattice symmetries but are metallic. 
However, if the transition out of the WMI occurs at intermediate correlation strength, it may allow for intermediate states -- compatible with recent experiments~\cite{tang2022dielectric,Zong2025_moSe2_quantum_melting} on quantum melting -- that respect lattice symmetries and are insulating.
We refer to these states as quantum charge liquids (QCLs) following Ref.~\cite{musser_et_al_2025}.
We specifically focus on gapped QCLs at fixed rational electronic filling $\nu = p/q$, where $p$ and $q$ are coprime integers. The Lieb-Schultz-Mattis-Oshikawa-Hastings (LSMOH) theorem \cite{Lieb1961_antiferromagnetic_chain, Oshikawa2000_commensurability, hastings_2004} constrains these gapped QCLs to exhibit fractional-charge excitations and topological order (TO). For 2d materials such as TMDs, additional constraints apply. In TMDs, the spin exchange scale is small \cite{Mak2022_semiconductor_moire}, and so spins are easily polarized by external magnetic fields. We therefore specialize to spinless fermions and also restrict to systems with time-reversal symmetry. Ref.\ \cite{musser_et_al_2025} further shows that that the minimal TO, i.e., TO of smallest ground state degeneracy on a torus, compatible with these constraints is $\mathbb{Z}_q$ ($\Z_{2q}$) for $q$ odd (even). 

Given these constraints on the allowed QCLs at filling $p/q$, a natural question arises: is there a direct defect-mediated quantum melting transition between the WMI and these minimal TOs? In this work, we find that the answer depends on the WMI's ordering pattern. On a 2d periodic lattice, the LSMOH theorem implies that the WMI must break lattice translation symmetry such that the unit cell of the ground state has size $nq$. From here on out, we denote the WMI with unit cell enlarged by a factor $nq$ as the $nq$-fold WMI. Moreover, by analogy with the definition of the minimal TO, we define the minimal WMI to be the $q$-fold WMI. Our main result is the following: a defect-mediated transition between the minimal fermionic WMI and the minimal fermionic TO is possible if and only if $q$ is odd. This result constrains the possible phase diagrams that include the minimal WMI. 

Our paper takes the approach of defect proliferation\footnote{The defects that we consider are confined in the WMI phase. Our focus is on approaching a quantum phase transition from the WMI where defect-antidefect pairs condense and give rise to the QCL. The term proliferation is used in this sense throughout the paper.}
to understand the gapped QCLs proximate to the minimal WMI. This approach to phase transitions has a rich history for describing melting, including but not limited to Refs.\ \cite{Kosterlitz_Thouless_1972_JPhysC, Halperin_Nelson_1978_PRL, Nelson_Halperin_1979_PRB}. More recently, this perspective was taken to study the quantum melting of crystals in Refs.\ \cite{Cho_et_al_2015,Pretko_Zhai_Radzihovsky_2019_PRB, Kumar_Potter_2019_PRB, Gromov_Radzihovsky_2024_RMP, matus2025defectswignercrystalsfractonelasticity}. 

This perspective has also been taken for WMIs specifically~\cite{balents_et_al_2005, balents_supplement, balents_triangular}.
A WMI with solely lattice translation symmetries has $q$ degenerate ground states, each related to each other by a lattice translation. If the minimal WMI has a unit cell of size $r\times s$, where $r$ and $s$ are positive integers that multiply to $q$, these ground states are labeled by a $\Z_r \times \Z_s$ order. Defects of the minimal WMI include both domain walls charged under the $\Z_r \times \Z_s$ order and junctions of domain walls.  When proliferated, the quantum numbers of these junctions restricts the allowed phases one can enter from the WMI. If the WMI is fermionic, additional restrictions arise. This perspective of defect proliferation was inspired by previous work on the 2d Neel-valence bond solid (VBS) transition in quantum magnets and the closely related superfluid-Mott transition. In particular, Ref.\ \cite{levin_senthil_2004} explains the paradigmatic VBS to Neel deconfined critical point through the perspective of defect proliferation from the VBS phase, and Ref.~\cite{balents_et_al_2005} uses a similar perspective to explain transitions out of the superfluid state on a 2d square lattice to a Mott insulating state. 

While our paper ultimately focuses on fermionic QCLs, we build off of previous work done on transitions into the gapped bosonic QCL. Examples of bosonic QCLs have been constructed in lattice models in Refs.\ \cite{Moessner_Raman_2008_QDM, Balents_Fisher_Girvin_2002_PRB, Senthil_Motrunich_2002_PRB, Motrunich_Senthil_2002_PRL, Motrunich_2003_PRB}. Additionally, field theoretic descriptions of gapped bosonic QCLs appear in Refs.\ \cite{Senthil_Fisher_2000_PRB, Balents_Nayak_Fisher_1999_PRB, balents_et_al_2005, Lannert_Fisher_Senthil_2001_PRB}. While these references primarily construct transitions from the superfluid state to the gapped bosonic QCLs, we present an alternate perspective, i.e. transitions from the bosonic WMI, in this work.

The paper is organized as follows. In Sec.\ \ref{sec:boson}, we first review the melting of a bosonic WMI at half filling on a square lattice following Ref.\ \cite{levin_senthil_2004}. We then discuss the accessible TOs for the minimal bosonic WMI for $q>2$ in the presence of only lattice translation symmetries. In Sec.\ \ref{sec:fermion}, we argue that the minimal fermionic WMI can melt via defect proliferation into the minimal TO if and only if $q$ is odd. In Sec.\ \ref{sec:QCL}, we take the alternate perspective and consider transitions out of the minimal TO. There, we show that the proximate fermionic WMI has $q$-fold ($2q$-fold) unit cell enlargement for $q$ odd (even). Finally, we conclude with a discussion of future directions in Sec.\ \ref{sec:discussion}.

\section{Quantum melting of bosonic WMIs}\label{sec:boson}
In this section, we describe the melting of the minimal bosonic WMIs at filling $p/q$ on a 2d lattice. As a brief review, we first discuss the melting of the filling $1/2$ bosonic WMI on a square lattice in the presence of lattice rotation symmetry as discussed in Ref.\ \cite{levin_senthil_2004}. We subsequently discuss the $q>2$ case in the presence of only lattice translation symmetry.

\subsection{Filling \texorpdfstring{$1/2$}{1/2} bosonic WMI in the presence of lattice rotation symmetry}
\begin{figure}
\centering
    \begin{tikzpicture}[scale = 0.5]
    \foreach \a in {0,1,2,3}
    {
        \foreach \i in {0,1,2,3,4}
        {
            \draw [dotted] (-0.5 + 8*\a, \i) -- (4.5 + 8*\a, \i);
            \draw [dotted] (\i + 8*\a, -0.5) -- (\i + 8*\a, 4.5);
        }
    }

    \foreach \i in {0,1,2,3,4}
    {
        \filldraw (0.5,\i) circle (3pt);
        \filldraw (2.5,\i) circle (3pt);

        \filldraw (8 + \i, 0.5) circle (3pt);
        \filldraw (8 + \i, 2.5) circle (3pt);
        
        \filldraw (17.5, \i) circle (3pt);
        \filldraw (19.5, \i) circle (3pt);

        \filldraw (24 + \i, 1.5) circle (3pt);
        \filldraw (24 + \i, 3.5) circle (3pt);
    }

    \draw[magenta] (-0.5, 1.5) -- (-0.5, 2.5) -- (1.5, 2.5) -- (1.5, 1.5) -- (-0.5,1.5);
    \draw[magenta] (8.5, 1.5) -- (8.5, 3.5) -- (9.5, 3.5) -- (9.5, 1.5) -- (8.5,1.5);
    \draw[magenta] (16.5, 1.5) -- (16.5, 2.5) -- (18.5, 2.5) -- (18.5, 1.5) -- (16.5,1.5);
    \draw[magenta] (24.5, 0.5) -- (24.5, 2.5) -- (25.5, 2.5) -- (25.5, 0.5) -- (24.5,0.5);
    
    \end{tikzpicture}
\caption{Caricatures of the four inequivalent stripe orders on the square lattice at $1/2$ filling. We draw the charge density wave pattern as being bond centered for convenience. Each pattern is related to the subsequent one by ${R}_{\pi/2}$. An example unit cell is drawn in magenta in each sketch.}\label{fig:bosonic_half}
\end{figure}
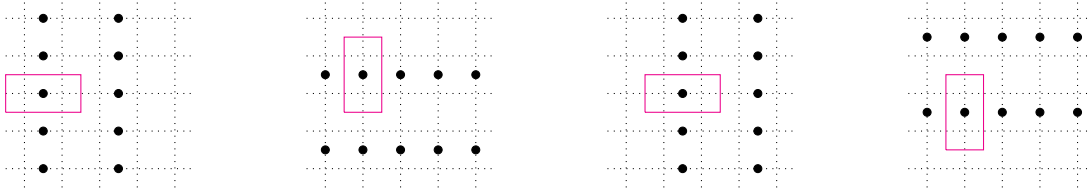

We first discuss the bosonic WMI at $1/2$ filling on the square lattice with lattice translation symmetries labeled $T_x$ and $T_y$, $\pi/2$ lattice rotation symmetry $R_{\pi/2}$ around the center of a unit cell plaquette, and time reversal symmetry $\mathcal{T}$. Such an insulator spontaneously breaks the lattice translation symmetries. We tune the lattice potential such that the lattice Hamiltonian favors the formation of stripe order (as opposed to checkerboard or plaquette order, both allowed in the presence of $R_{\pi/2}$). In the added presence of rotation symmetry, the Hamiltonian has four degenerate stripe-ordered ground states: two inequivalent horizontal stripe patterns related by $T_y$ and two inequivalent vertical stripe patterns related by $T_x$. We display caricatures of the four stripe orders in Fig.\ \ref{fig:bosonic_half}. Note that while we represent the bosons as living on the links of the square lattice, as in a dimer model, the true density wave pattern may be smoothed out and is model-specific. These stripe ordered states can melt via proliferation of defects and form a superfluid or a bosonic $\mathbb{Z}_2$ topological order (TO) as we review below. The defect proliferation picture is discussed in more detail in Ref.\ \cite{levin_senthil_2004} for the valence bond solid to Neel transition. The above transitions are also described in Refs.\ \cite{balents_et_al_2005, senthil_et_al_2004, Sachdev_review_2008, senthil_balents_sachdev_vishwanath_fisher_2004}. 

\begin{figure}
\centering
\begin{tikzpicture}[scale = 0.5]
\foreach \x in {-4,-3,-2,-1,0,1,2,3,4}
{
    \draw [dotted] (-4.5,\x) -- (4.5, \x);
    \draw [dotted] (\x, -4.5) -- (\x, 4.5);
}

\foreach \i in {0,1,2,3,4}
{
    \filldraw (-1.5,\i) circle (3pt);
    \filldraw (-3.5,\i) circle (3pt);
    \filldraw (\i, 1.5) circle (3pt);
     \filldraw (\i, 3.5) circle (3pt);
    \filldraw (1.5, -\i) circle (3pt);
    \filldraw (3.5, -\i) circle (3pt);
     \filldraw (-\i, -1.5) circle (3pt);
     \filldraw (-\i, -3.5) circle (3pt);
}

\draw[blue] (-0.5,4.5) -- (-0.5,0.5);
\draw[blue] (0.5,-4.5) -- (0.5,-0.5);
\draw [blue,domain=180:270] plot ({0.5*cos(\x)}, {0.5*sin(\x) + 0.5});
\draw [blue,domain=0:90] plot ({0.5*cos(\x)}, {0.5*sin(\x) - 0.5});

\draw[blue] (-4.5,-0.5) -- (-0.5,-0.5);
\draw[blue] (4.5,0.5) -- (0.5,0.5);
\draw [blue,domain=90:180] plot ({0.5*cos(\x)+0.5}, {0.5*sin(\x)});
\draw [blue,domain=270:360] plot ({0.5*cos(\x)-0.5}, {0.5*sin(\x)});
\draw[magenta] (-0.5,0.5) -- (0.5,0.5) -- (0.5,-0.5) -- (-0.5,-0.5) -- (-0.5,0.5);

\end{tikzpicture} \hspace{0.15\textwidth}
        \begin{tikzpicture}[scale = 0.5]
\foreach \x in {-4,-3,-2,-1,0,1,2,3,4}
{
    \draw [dotted] (-4.5,\x) -- (4.5, \x);
    \draw [dotted] (\x, -4.5) -- (\x, 4.5);
}

\foreach \i in {0,1,2,3,4}
{
    \filldraw (-0.5,\i) circle (3pt);
    
    \filldraw (-2.5,\i) circle (3pt);
}
\foreach \i in {0,1,2,3}
{
    \filldraw (\i + 1, 1.5) circle (3pt);
     \filldraw (\i + 1, 3.5) circle (3pt);
}
\foreach \i in {0,1,2,3,4}
{
    \filldraw (2.5, -\i) circle (3pt);
}
\foreach \i in {0,1,2,3,4,5}
{
     \filldraw (-\i + 1, -1.5) circle (3pt);
     \filldraw (-\i + 1, -3.5) circle (3pt);
}

\draw[blue] (-0.5 + 1,4.5) -- (-0.5 + 1,0.5);
\draw[blue] (0.5 + 1,-4.5) -- (0.5 + 1,-0.5);
\draw [blue,domain=180:270] plot ({0.5*cos(\x) + 1}, {0.5*sin(\x) + 0.5});
\draw [blue,domain=0:90] plot ({0.5*cos(\x) + 1}, {0.5*sin(\x) - 0.5});

\draw[blue] (-4.5,-0.5) -- (-0.5 + 1,-0.5);
\draw[blue] (4.5,0.5) -- (0.5 + 1,0.5);
\draw [blue,domain=90:180] plot ({0.5*cos(\x)+0.5 + 1}, {0.5*sin(\x)});
\draw [blue,domain=270:360] plot ({0.5*cos(\x)-0.5 + 1}, {0.5*sin(\x)});
\draw[magenta] (-0.5 + 1,0.5) -- (0.5 + 1,0.5) -- (0.5 + 1,-0.5) -- (-0.5 + 1,-0.5) -- (-0.5 + 1,0.5);

\end{tikzpicture} 
\caption{Caricatures of the two inequivalent vortex configurations on the square lattice at 1/2 filling. The blue lines are domain walls. The right picture is obtained from the left picture by acting with $T_x$. Note that the right vortex has the same stripe order in quadrants 1 and 3 but opposite stripe orders in quadrants 2 and 4. Thus, the right vortex has opposite vorticity under $\mathbb{Z}_4$. Each vortex has an empty half unit cell at the vortex core.}\label{fig:bosonic_half_vortices}
\end{figure}
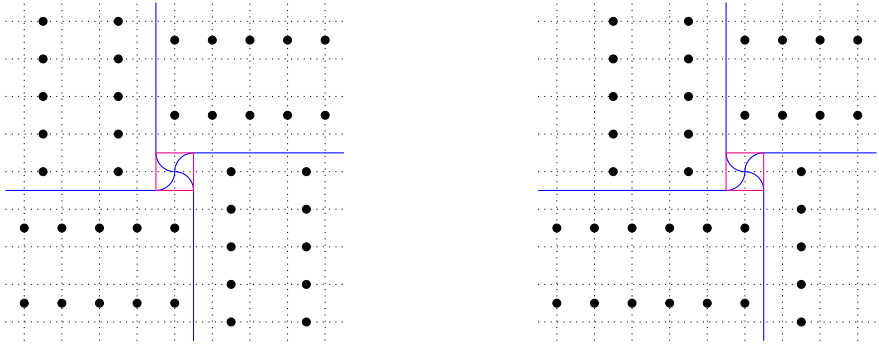

We discuss two elementary defects for the square lattice at filling $1/2$: domain walls and vortices. A domain wall is a defect that separates two different stripe orderings. As in Ref.\ \cite{balents_et_al_2005}, we define domain walls as being insulating (excitation spectrum is gapped in the presence of a domain wall) and charge neutral (no excess charge per unit cell).\footnote{Charged domain walls can be obtained from composites of charge neutral domain walls and vortices/domain wall junctions.} The other distinct elementary defect we discuss, a vortex, consists of a vortex core and 4 domain walls that terminate at the core. Note that all four stripe order patterns are related by $R_{\pi/2}$. Thus, we encode the stripe order in a $\mathbb{Z}_4$ order parameter, and the elementary vortex corresponds to a $\mathbb{Z}_4$ vortex of the stripe order parameter. The center of the $\mathbb{Z}_4$ vortex has an extra unoccupied half unit cell, and so the $\mathbb{Z}_4$ vortex carries a total excess charge $-1/2$. Under $T_x$, $T_y$, or $R_{\pi/2}$, the vortex transforms to another vortex flavor with the same charge but opposite vorticity under $\mathbb{Z}_4$. We display caricatures of both vortices in Fig.\ \ref{fig:bosonic_half_vortices}. Thus, inserting both flavor of vortices at a single unit cell corresponds a defect with an empty unit cell and no net vorticity, i.e., the removal of a boson from the stripe ordered state without changing the stripe order parameter. As a result, we can identify the product of both vortices with the physical anti-boson, and so proliferating both flavors of vortices and antivortices disorders the stripe ordered bosonic WMI and creates a superfluid.

We describe the field theory of the transition with a parton mean field description. We denote the fields for the two flavors of vortices as $\phi_+$ and $\phi_-$ for positive and negative vorticity. The flavors transform into each other under $T_x$, $T_y$, and $R_{\pi/2}$. We denote the physical boson as $\Phi$. Because the product of $\phi_+$ and $\phi_-$ is the physical antiboson $\Phi^\dagger$, we expect the following Lagrangian to describe the transition between the stripe ordered bosonic WMI and superfluid:
\begin{equation}
    \mathcal{L} = \mathcal{L}[\phi_+,a_+] + \mathcal{L}[\phi_-, a_-] - V(\phi_+, \phi_-) + \frac{1}{2\pi} b \wedge \dd(a_+ + a_- + A) + \mathcal{L}_{\text{mp}}, \label{eq:lagrangian_rot}
\end{equation}
where $a_+$ and $a_-$ are internal $U(1)$ gauge fields that are also permuted by $T_x$, $T_y$, and $R_{\pi/2}$, $b$ is a Lagrange multiplier 1-form, $A$ is the background $U(1)$ gauge field under which the physical boson is charged, $V$ is a potential that respects the lattice symmetries (namely, interchange of $\phi_+$ and $\phi_-$), 
\begin{equation}
    \mathcal{L}[\phi,a] = |(\partial_\mu - ia_\mu) \phi|^2,
\end{equation}
and $\mathcal{L}_\text{mp}$ is the contribution of monopoles to the Lagrangian. 
Generically, $V$ takes the form 
\begin{equation}
    V(\phi_+, \phi_-) = m_\phi^2(|\phi_+|^2 + |\phi_-|^2) + \lambda(|\phi_+|^2 + |\phi_-|^2)^2 + \tilde{\lambda} |\phi_+|^2 |\phi_-|^2 + \cdots,
\end{equation}
where $m_\phi^2$, $\lambda$ and $\tilde{\lambda}$ are real numbers.
To fix the form of $\mathcal{L}_\text{mp}$, we first note that terms of the form $\mathcal{M}_+\mathcal{M}_-^\dagger$ are generically allowed in $\mathcal{L}_\text{mp}$ by lattice translation, and so we can identify the two monopoles with each other. Furthermore, note that in the translationally invariant phase each unit cell carries charge $1/2$ of the background $U(1)$ gauge field, and by the Lagrange multiplier constraint and translation symmetries, each unit cell must also carry charge $-1/2$ of the internal gauge fields $a_+$ and $a_-$.  Thus, the monopoles $\mathcal{M}_\pm$ of $a_\pm$ transform projectively under lattice translation and rotation. In other words, the action of the lattice translation and rotation symmetries on the monopoles obey the following projective symmetry group (PSG) algebraic relations:
\begin{align}
    \begin{split}
        T_x T_y &= -T_y T_x\\
        T_x R_{\pi/2}&=  R_{\pi/2} T_y^{-1}\\
         T_y R_{\pi/2}&=  R_{\pi/2} T_x\\
         (R_{\pi/2})^4 &= 1. 
    \end{split}
\end{align}
For $\mathcal{M}$ to realize the PSG relations, it should be thought of as a $2$ component vector $\mathcal{M}_\ell$. Moreover, because $T_x$, $T_y$, and $R$ permute $a_\pm$, they also permute the monopoles $\mathcal{M}_\pm$. A (nonunique) choice of PSG action on $\mathcal{M}_\pm$ is 
\begin{equation}
    T_x = \sigma^y, \quad T_y = \sigma^x, \quad R_{\pi/2} = \frac{1}{\sqrt{2}}(\sigma^x + \sigma^y),
\end{equation}
where $\sigma^x$ and $\sigma^y$ are the Pauli matrices. This PSG action fixes the lowest order nontrivial monopole vertex term (i.e., an operator that creates/annihilates at least one flavor of monopole) to be 
\begin{equation}
    \mathcal{L}_\text{mp} =\gamma \Re{(\mathcal{M}_+ \mathcal{M}_-^\dagger)^4}, \label{eq:monopole_rot}
\end{equation}
where $\gamma>0$. Note that performing standard particle-vortex duality to Eq.\ \eqref{eq:lagrangian_rot} yields the dual vortex theory discussed in Ref.\ \cite{balents_et_al_2005}. 

In the bosonic stripe ordered phase, both $\phi_+$ and $\phi_-$ are gapped (i.e., $m_\phi^2 > 0$), and the monopoles  $\mathcal{M}_\pm$ are condensed. Moreover, $\mathcal{L}_\text{mp}$ fixes $\mathcal{M}_+ = e^{in\pi/2} \mathcal{M}_-$ in the monopole condensed phase, where $n$ labels the $\mathbb{Z}_4$ clock order parameter. The insulator-superfluid transition occurs when $V(\phi_+,\phi_-)$ is tuned such that both $\phi_+$ and $\phi_-$ condense with equal magnitude. 

\begin{figure}
\centering
    \begin{tikzpicture}[scale = 0.5]
\foreach \x in {-4,-3,-2,-1,0,1,2,3,4}
{
    \draw [dotted] (-4.5,\x) -- (4.5, \x);
    \draw [dotted] (\x, -4.5) -- (\x, 4.5);
}

\foreach \i in {0,1,2,3,4}
{
    \filldraw (-1.5,\i) circle (3pt);
    \filldraw (-3.5,\i) circle (3pt);
}
\foreach \i in {1,2,3,4}
{
    \filldraw (1.5,\i) circle (3pt);
    \filldraw (3.5,\i) circle (3pt);
}
\foreach \i in {1,2,3,4}
{
    \filldraw (-0.5,-\i) circle (3pt);
    \filldraw (-2.5,-\i) circle (3pt);
}
\filldraw (0, 1.5) circle (3pt);
\filldraw (0, 3.5) circle (3pt); 
\filldraw (1, -1.5) circle (3pt);
\filldraw (1, -3.5) circle (3pt); 
\filldraw (0.5, 0) circle (3pt); 
\foreach \i in {0,1,2,3,4}
{
    \filldraw (2.5, -\i) circle (3pt);
}
\draw[blue] (-0.5,4.5) -- (-0.5,0.5);
\draw[blue] (0.5,-4.5) -- (0.5,-0.5);

\draw[blue] (-4.5,-0.5) -- (-0.5,-0.5);
\draw[blue] (4.5,0.5) -- (0.5,0.5);

\draw[blue] (-0.5 + 1,4.5) -- (-0.5 + 1,0.5);
\draw[blue] (0.5 + 1,-4.5) -- (0.5 + 1,-0.5);

\draw[blue] (-4.5,-0.5) -- (-0.5 + 1,-0.5);
\draw[blue] (4.5,0.5) -- (0.5 + 1,0.5);
\draw[magenta] (0.5,-0.5) -- (-0.5,-0.5) --  (-0.5,0.5) -- (0.5,0.5)  ;
\draw[magenta] (-0.5 + 1,0.5) -- (0.5 + 1,0.5) -- (0.5 + 1,-0.5) -- (-0.5 + 1,-0.5) ;

\end{tikzpicture}
  
\caption{Caricature of the defect configuration $\phi_+^\dagger \phi_-$, dressed with the appropriate Wilson lines, at 1/2 filling that, when proliferated, results in a $\mathbb{Z}_2$ TO. The defect has vorticity $4\pi$ and has a core consisting of a unit cell.}\label{fig:bosonic_half_TO}
\end{figure}
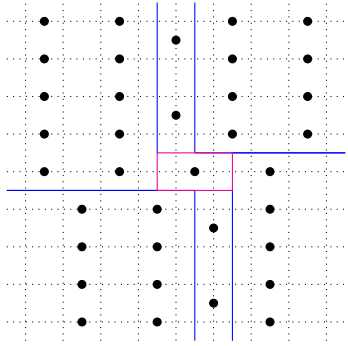

From the field theory, we also see a path to the $\mathbb{Z}_2$ TO: the condensation of $\phi_+^\dagger \phi_-$. Physically, this field is charge neutral, has $4\pi$ vorticity, and transforms to its Hermitian conjugate under $T_x$, $T_y$, or $R$. Therefore, we expect that condensing this field results in the $\mathbb{Z}_2$ TO and restores the lattice symmetries. We depict a caricature of this defect in Fig.\ \ref{fig:bosonic_half_TO}. In the field theory perspective, condensing $\phi_+^\dagger \phi_-$ results in the following system of equations for the two internal gauge fields:
\begin{align}
\begin{split}
    a_+ - a_- &\equiv 0 \mod 2\pi \\
    a_+ + a_- + A &\equiv 0\mod 2\pi.
\end{split}
\end{align}
Thus, we find that $2a_+ + A \equiv 0 \mod 2\pi$, i.e.\ $a_+$ is Higgsed from $U(1)$ to $\mathbb{Z}_2$. Moreover, the $\mathbb{Z}_2$ TO is $U(1)$ enriched: the $e$-particle of the TO, $\phi_+$, has charge $-1/2$ under the background $U(1)$ gauge field. 

\subsection{Minimal bosonic WMI for \texorpdfstring{$q>2$}{q>2} in the absence of lattice rotation symmetry}\label{sec:bosonic_q_greater_2}

We now discuss the minimal bosonic WMI at $p/q$ filling, $q>2$ on a 2d lattice with lattice translation symmetries labeled $T_x$ and $T_y$, time reversal symmetry $\mathcal{T}$, but no lattice rotation symmetry. We assume that the size of the unit cell for this WMI is $r\times s$, where $rs = q$. For general $q$, in the absence of extra symmetries such as lattice rotation, we cannot uniquely fix the order parameter patterns of the defects as we did in the previous section. Nevertheless, we can describe the melting transitions from the bosonic WMI into the superfluid and the topological order from the field theory itself.

We follow Ref.\ \cite{balents_et_al_2005} for the rest of this section and describe the superfluid to WMI transition via a theory of the superfluid vortices. At filling $p/q$, the superfluid vortices experience a dual magnetic field with flux $2\pi p/q$ in each unit cell. As a result, the vortex fields transform projectively under lattice translation, i.e., $T_x T_y T_x^{-1} T_y^{-1} = \omega$, where $\omega = \exp{2\pi i p/q}$. Upon diagonalizing the vortex Hamiltonian, we find $q$ distinct degenerate vortex minima, as in the Hofstadter problem. As a result, there are $q$ distinct species of vortices, labeled $\varphi_\ell$ where $\ell$ ranges from $0$ to $q-1$.

\subsubsection{Vertical stripe WMIs}\label{sec:vertical_stripe_WMIs}
We first consider the case where the unit cell has size $q \times 1$. We choose a basis of $\varphi_\ell$ such that lattice translations act as
\begin{equation}
    T_x: \varphi_\ell \to \varphi_{\ell+1}, \quad T_y: \varphi_{\ell} \to \omega^{-\ell}\varphi_{\ell - 1} \label{eq:q_odd_psg}
\end{equation}
for $q$ odd and 
\begin{equation}
    T_x: \varphi_\ell \to \varphi_{\ell+1}, \quad T_y: \varphi_{\ell} \to \omega^{-\ell + 1/2}\varphi_{\ell - 1}\label{eq:q_even_psg}
\end{equation}
for $q$ even.
Here, addition and subtraction in the index are taken modulo $q$. We explicitly show how to construct this basis from the vortex description in Ref.\ \cite{balents_et_al_2005} in App.\ \ref{app:basis_choice}. This choice of basis is convenient but not unique. In principle, for a convenient description of melting transitions out of the bosonic WMI, we may choose any basis such that condensing a locked pattern of all $\varphi_\ell$ fields results in the vertical stripe WMI. For this choice of basis, in the particle-vortex dual description, the WMI phase is written in terms of $q$ gapped confined parton excitations $\phi_\ell$ that are dual to the $\varphi_\ell$ fields and become deconfined at the critical point. We discuss the convenience of basis further in App.\ \ref{app:explanation}.

The Lagrangian description of the vortex theory is 
\begin{equation}
   \mathcal{L} = \sum_\ell \mathcal{L}[\varphi_\ell, b] + V_v(\varphi_\ell) +  \frac{p}{2\pi} b \wedge \dd A,
\end{equation}
where $b$ is an internal $U(1)$ gauge field, and $V_v(\varphi_\ell)$ is a vortex potential that respects the PSG translation symmetries. In the superfluid phase, all vortices are gapped and the theory is translation symmetric. In the bosonic WMI phase, a vector of $\varphi_\ell$ fields condenses, i.e., all $\varphi_\ell$ fields condense and are locked to each other. A WMI with a $q\times 1$ unit cell requires that the system be invariant only under $T_x^q$ and $T_y$ (i.e., we should have $q$ $\varphi_\ell$ condensation patterns that permute under $T_x$ and are eigenvalues of $T_y$). By inspection, we find that the $\varphi_\ell$ condensation pattern that satisfies this constraint for $q$ odd is 
\begin{equation}
    \ev{\varphi_\ell} = \omega^{-(\ell - n) (\ell - n + 1)/2} v_0,
\end{equation}
where $n$ labels the $\mathbb{Z}_q$ valued stripe order parameter and $v_0$ is some nonzero complex number.
For $q$ even, the condensation pattern is 
\begin{equation}
    \ev{\varphi_\ell} = \omega^{-(\ell -n)^2/2} v_0,
\end{equation}
Note that acting with $T_x$ shifts $n$ by $1$. The order parameter that detects the stripe ordered phase (i.e., the value of $n$) for $q$ odd is 
\begin{equation}
    B_\text{odd}= \sum_{\ell=0}^{q-1} \omega^{-\ell} \varphi_{\ell + 1}^\dagger \varphi_\ell = \varphi_\ell^\dagger D_{\text{odd, }\ell \ell'} \varphi_{\ell'},
\end{equation}
where in the second line, we represent the sum with the matrix $D_{\text{odd, }\ell \ell'}$ for later convenience.
Likewise, the order parameter that detects the stripe ordered phase for $q$ even is 
\begin{equation}
    B_\text{even}= \sum_{\ell=0}^{q-1} \omega^{-\ell -1/2} \varphi_{\ell + 1}^\dagger \varphi_\ell = \varphi_\ell^\dagger D_{\text{even, }\ell \ell'} \varphi_{\ell'},
\end{equation}
Both order parameters take values $\omega^{-n}$. Thus, to ensure that the bosonic WMI has horizontal stripes, we include in the potential $V_v$ the translation symmetric term $-\lambda \Re{B^q}$, where $\lambda$ is some positive real number.

To discuss the defects that must be proliferated to enter the superfluid or the topologically ordered state, we apply the standard particle-vortex duality to our Lagrangian:
\begin{equation}
    \mathcal{L}= \sum_\ell \left(\mathcal{L}[\phi_\ell, a_\ell]  + \frac{1}{2\pi} b \wedge d a_\ell \right)+ V(\phi_\ell) + \mathcal{L}_\text{mp} +  \frac{p}{2\pi} b \wedge d A,
    \label{eq:dualWMI}
\end{equation}
where $\phi_\ell$ is dual to $\varphi_\ell$ and couples to internal gauge fields $a_\ell$, $V(\phi_\ell)$ is a potential for the dual fields, and we supplement our Lagrangian with monopole terms in $\mathcal{L}_\text{mp}$. In the superfluid phase, all $\phi_\ell$ fields condense with equal magnitude. In the bosonic WMI phase, all $\phi_\ell$ fields are gapped. 

First, note that the physical boson $\Phi$, upon integrating out the Lagrange multiplier $b$, satisfies $(\Phi^\dagger)^p = \phi_0 \ldots \phi_{q-1}$. Because $\phi_\ell$ creates vortices of $\varphi_\ell$, and $T_x$ and $T_y$ permute $\varphi_\ell$ up to a phase, we expect that $T_x$ and $T_y$ also permute $\phi_\ell$ up to a (different) phase. Moreover, note that $\phi_\ell$ does not see a magnetic field in every unit cell and thus, translation does not act projectively on $\phi_\ell$. Thus, we can choose a gauge in which 
\begin{align}
\begin{split}
    T_x: \phi_\ell \to \phi_{\ell + 1}&, \quad T_y: \phi_{\ell} \to \phi_{\ell - 1}, \\
     T_y: a_\ell \to a_{\ell + 1}&, \quad T_y: a_{\ell} \to a_{\ell - 1}, 
\end{split}
\end{align}
Then, $V(\phi_\ell)$ is a potential that is invariant under permutation of $\phi_\ell$. To fix the monopole terms in the action, first note that the monopole $\mathcal{M}_{\ell}$ of $a_\ell$ also sees an effective $2\pi p/q$ magnetic flux. Thus, the $\mathcal{M}_{\ell}$ fields transform projectively under $T_x$ and $T_y$. Moreover, under particle-vortex duality, $\varphi_\ell$ and $M_\ell$ must have the same transformation properties under $T_x$ and $T_y$. 
Thus, $\mathcal{L}_\text{mp}$ takes the form
\begin{equation}
    \mathcal{L}_\text{mp} = -\lambda \Re{\left(\mathcal{M}_\ell^\dagger D_{\ell,\ell'} \mathcal{M}_{\ell'}\right)^q} + \cdots,
\end{equation}
where here $D$ is either $D_\text{odd}$ or $D_\text{even}$ depending on the value of $q$.

Thus, we have fixed all the terms in the dual Lagrangian in Eq.\ \eqref{eq:dualWMI}. Upon integrating out $b$, the path to the superfluid and the topologically ordered state is apparent. Higgsing all $\phi_\ell$ with equal amplitude results in the superfluid state. To arrive at the $\mathbb{Z}_q$ TO, we instead Higgs all $\phi_{\ell}^\dagger \phi_{\ell+1}$  with equal magnitude.\footnote{If we approach this confinement-deconfinement phase transition from the QCL side, it is described as a condensation of the $q$-species of the $m$-particles, $\vphi_l$, of the $\mathbb{Z}_q$ TO (see Sec. \ref{sec:QCL}). The $m$-particles can be described by a real vector field of dimension $2q$. The transition, modulo fractionalization, then has the same universality as an $O(2q)$ model with anisotropies that are generically allowed by the lattice translation symmetries. For $q=2$ however, the vector field is two-dimensional because $\vphi_l^2$ is condensed and the fields can be chosen to be real. In this case, the anisotropy is irrelevant and hence the transition is in the XY* universality class~\cite{carmona_2000,senthil_dqcp_review}. }
Such a condensate results in the following system of equations for the $q$ internal gauge fields:
\begin{align}
\begin{split}
    a_{\ell}- a_{\ell'} &\equiv 0 \mod 2\pi \\
    \sum_{\ell=0}^{q-1} a_{\ell} + pA &\equiv 0\mod 2\pi. \label{eq:zqto}
\end{split}
\end{align}
Thus, we find that $qa_{0} + pA \equiv 0 \mod 2\pi$, i.e.\ $a_0$ is Higgsed from $U(1)$ to $\mathbb{Z}_q$. Moreover, as we show below, the $\mathbb{Z}_q$ TO is $U(1)$ enriched. Upon condensing $\phi_{\ell}^\dagger \phi_{\ell+1}$, all the $\phi_{\ell}$ fields are identified with each other, and so we arrive at the effective Lagrangian
\begin{equation}
    \mathcal{L}_\text{TO} = L[\phi_{0}, a_{0}] + \frac{1}{2\pi} b \wedge \dd (qa_{0} + pA).
\end{equation}
Thus, $\phi_{0}$ now carries charge $-p/q$ under the background $U(1)$ gauge field and is thus identified with the $e$ particle of the TO.

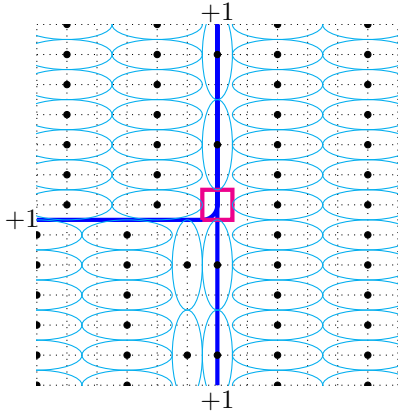
\begin{figure}
\centering
\begin{tikzpicture}[scale=0.4]
  \node at (0,6.4) {$+1$};
    \node at (0,-6.5) {$+1$};
    \node at (-6.5,-0.5) {$+1$};
\clip(-6,-6) rectangle (6,6);
\foreach \i in {-6,...,6} {
    \draw[dotted] (\i,-6) -- (\i,6);
    \draw[dotted] (-6,\i) -- (6,\i);
  }

  \draw[blue,ultra thick] (-6,-0.5) -- (-0.5,-0.5)to [out=0,in=270] (0,0) -- (0,6) -- (0,-6);

\draw[magenta,ultra thick](-0.5,-0.5) -- (-0.5,0.5) -- (0.5,0.5) -- (0.5,-0.5) -- (-0.5,-0.5);

\foreach \i in {0,...,6} {
\filldraw (-2,\i) circle (3pt);
\draw[cyan](-2,\i)circle[x radius = 1.5, y radius = 0.5];
\filldraw (-5,\i) circle (3pt);
\draw[cyan](-5,\i)circle[x radius = 1.5, y radius = 0.5];
  }

\foreach \i in {1,...,6} {
\filldraw (-3,-\i) circle (3pt);
\draw[cyan](-3,-\i)circle[x radius = 1.5, y radius = 0.5];
\filldraw (-6,-\i) circle (3pt);
\draw[cyan](-6,-\i)circle[x radius = 1.5, y radius = 0.5];
}

\foreach \i in {-6,...,6} {
\filldraw (2,\i) circle (3pt);
\draw[cyan](2,\i)circle[x radius = 1.5, y radius = 0.5];
\filldraw (5,\i) circle (3pt);
\draw[cyan](5,\i)circle[x radius = 1.5, y radius = 0.5];
}
\filldraw (0,2) circle (3pt);
\draw[cyan](0,2)circle[x radius = 0.5, y radius = 1.5];
\filldraw (0,5) circle (3pt);
\draw[cyan](0,5)circle[x radius = 0.5, y radius = 1.5];
\filldraw (0,-2) circle (3pt);
\draw[cyan](0,-2)circle[x radius = 0.5, y radius = 1.5];
\filldraw (-1,-2) circle (3pt);
\draw[cyan](-1,-2)circle[x radius = 0.5, y radius = 1.5];
\filldraw (0,-5) circle (3pt);
\draw[cyan](0,-5)circle[x radius = 0.5, y radius = 1.5];
\filldraw (-1,-5) circle (3pt);
\draw[cyan](-1,-5)circle[x radius = 0.5, y radius = 1.5];

\end{tikzpicture}

\caption{A caricature of the defect $\phi_\ell(x) \exp{-i\int^x a_\ell}$, dressed by an operator such that each domain has vertical stripes, for the trimer model at filling $1/3$. The domain walls are blue, with charge $+1$ under the stripe order parameter group $\mathbb{Z}_3$, and the junction center is an empty magenta box. We represent the bosons as living in trimers, drawn in cyan, to show explicitly the empty site at the junction center. Some vertical trimers are necessary along domain walls to ensure that the filling fraction remains $1/3$ outside of the junction center. As before, the charge density wave pattern and shapes of the domain walls are model specific.}\label{fig:bosonic_third_vortices}
\end{figure}

What physically does the defect $\phi_\ell$ look like? Recall that the stripe ordered state has the following $\varphi_\ell$ condensation pattern for $q$ odd (and a similar one for $q$ even):
\begin{equation}
(\ev{\varphi_0}, \ldots, \ev{\varphi_\ell}, \ldots \ev{\varphi_{q-1}}) = (\ev{\varphi_0}, \ldots, \omega^{-(\ell - n) (\ell - n + 1)/2} \ev{\varphi_0}, \ldots, \omega^{-(q - n-1) (q-n)/2} \ev{\varphi_0}).
\end{equation}
Because $\phi_\ell$ is the vortex creation operator of $\varphi_\ell$, it winds the phase of $\varphi_\ell$ around its vortex center while leaving the other $\varphi_{\ell'}$s invariant. Thus, applying $\phi_\ell$ creates domains of different stripe configurations around what we call a domain wall junction -- these stripe configurations may have vertical, horizontal, or diagonal stripes.\footnote{For example, the three vertical stripe condensates for filling $1/3$ are $\ev{\vec{\varphi}} \sim (1, 1, \omega^2) v_0$, $(1,\omega^2, 1)v_0$, and $(\omega^2, 1, 1)v_0$ for some complex number $v_0$. A $2\pi$ vortex in one of these entries does not result in domains of vertical stripes.} However, recall that $V(\varphi_\ell)$ energetically favors configurations with vertical stripes. Therefore, the energetically favorable defect operator is not $\phi_\ell$ but is $\phi_\ell$ dressed with domain wall operators to ensure that all defect domains have vertical stripes. Because this dressing is model specific, we do not comment on it further. Moreover, note that because $T_x$ permutes both $\phi_\ell$ and $a_\ell$ and because $\phi_\ell$ are partons of the physical boson $\Phi$, the gauge invariant operator $\phi_\ell(x) \exp{-i\int^x a_\ell}$ must carry physical charge $-p/q$ by translation symmetry.
We display an example of the domain wall junction $\phi_\ell(x) \exp{-i\int^x a_\ell}$ in Fig.\ \ref{fig:bosonic_third_vortices} for a trimer model, where the defect center is an empty lattice site. The general pictorial form of $\phi_\ell(x) \exp{-i\int^x a_\ell}$ is model specific. In particular, we stress that the size of the domains and the number of domain walls surounding the junction are not universal. 
We similarly can ask what physical defect $\phi_{\ell}(x)^\dagger \phi_{\ell'}(x)\exp{-i\int^x (a_{\ell'}-a_\ell)}$ corresponds to, i.e., what physical defect must be proliferated to enter the topological order. This defect carries no physical charge but may still disorder the $\mathbb{Z}_q$ order paramter. Thus, we identify this defect with a configuration of loops of uncharged domain walls of the $\mathbb{Z}_q$ order parameter. 

We finally note that further fractionalization of the boson into defects is generically allowed. For example, an extra $2$-fold fractionalization would allow a transition from the $q$-fold bosonic WMI into the $\mathbb{Z}_{2q}$ TO.\footnote{Coming from the TO at filling $p/q$, we can view each unit cell as having $e^2$, where $e$ is the $e$ particle of the $\mathbb{Z}_{2q}$ TO carrying charge $p/(2q)$ due to the further fractionalization. Then, condensing the $m$ particle would create a $q$-fold bosonic WMI per the arguments in Sec.\ \ref{sec:QCL}.} This type of transition can occur in lattice models where the unit cell basis has two sites (see e.g.\ Ref.\ \cite{Motrunich_Senthil_2002_PRL}). However, in this work, we focus on the simpler case of quantum melting transitions into the minimal allowed TO.

\subsubsection{General bosonic WMIs}
To carry over the arguments from the previous section, it suffices to find a basis of superfluid vortices such that condensing a locked pattern of all superfluid vortices results in a WMI with $r\times s$ size unit cells, as we discuss in App. \ref{app:explanation}.

First, we note that if $s=q$ and $r=1$, the basis from Ref.\ \cite{balents_et_al_2005} suffices. In this basis, which we label $\tilde{\varphi}_\ell$, the action of lattice translation is
\begin{equation}
    T_x: \tilde{\varphi}_\ell \to \tilde{\varphi}_{\ell+1}, \quad T_y: \tilde{\varphi}_\ell \to \omega^{-\ell} \tilde{\varphi}_\ell.
\end{equation}
To obtain $1\times q$ unit cells, i.e. horizontal stripes, the condensate pattern is 
\begin{equation}
    \ev{\tilde{\varphi}_\ell} = \omega^{n\ell} v_0,
\end{equation}
where $n$ labels the $\mathbb{Z}_q$ valued stripe order parameter and $v_0$ is some nonzero complex number.

For $r$ and $s$ both greater than $1$, we can always perform an arbitrary unitary rotation $U$ of the $\tilde{\varphi}_\ell$ fields to arrive at a basis $\varphi_\ell'$ where condensing a vector of $\varphi_\ell'$ with all nonzero entries results in a symmetry breaking pattern with $r\times s$ unit cells. After this unitary rotation, the lattice translations $T_x$ and $T_y$ are modified to $UT_xU^{-1}$ and $UT_y U^{-1}$. These new operators, while they satisfy the PSG relations, are generically not simple clock or shift operators in the $\varphi_\ell'$ basis. In particular, we can find a basis in which $T_x$ is a shift operator but $T_y$ no longer is a simple operator as discussed in App.\ \ref{app:basis_choice}. As a result, under particle vortex duality, the resulting bosonic partons $\phi_\ell'$ that are dual to $\varphi_\ell'$ permute under $T_x$ but transform nontrivially under $T_y$. Nevertheless, we still can interpret $\phi_\ell'$ as a domain wall junction, and we still can describe the transition into the superfluid as the simultaneous Higgsing of the $q$ $\phi_\ell'$ fields and the transition into the topological order as the simulatenous Higgsing of the bilinears $(\phi_{\ell}')^\dagger \phi_{\ell+1}'$.

\section{Quantum melting of minimal fermionic WMIs} \label{sec:fermion}
We now discuss the quantum melting of the minimal fermionic WMIs into QCLs on the square lattice at filling $p/q$. We specifically consider systems symmetric under $G = \mathbb{Z}^2\times [U(1)_f \rtimes \mathbb{Z}_2^T]$, where $\mathbb{Z}^2$ encodes the lattice translation symmetries, $U(1)_f$ encodes charge conservation with the restriction that the charge modulo $2$ is the fermion parity, and $\mathbb{Z}_2^T$ encodes the order 2 time-reversal symmetry that does not change the $U(1)$ charge. We also assume take the electrons to be a Kramers singlet, i.e. time reversal squares to the identity for the electrons. Under this symmetry, we find that the melting story differs for odd and even $q$. 

\subsection{Odd \texorpdfstring{$q$}{q}} \label{sec:odd_q}
We first consider odd $q$ and consider a minimal fermionic WMI with unit cell size $r\times s$. Following the discussion of Sec.\ \ref{sec:bosonic_q_greater_2}, we describe the transition as the Higgsing of $q$ elementary domain wall junctions related to each other by $T_x$. Because they are related to each other by translation, they must have the same self-statistics. We consider two cases, even $p$ and odd $p$.

\subsubsection*{Odd \texorpdfstring{$p$}{p}}
For odd $p$, the product of the junctions yields $p$ copies of the local anti-fermion, and because $q$ is odd, each junction must be fermionic. If we label the annihilation operator for each junction by $f_{\ell}$, where  $\ell$ ranges from $0$ to $q-1$, we have the relation $(c^\dagger)^p = \prod_{\ell} f_{\ell}$, where $c^\dagger$ is the creation operator of the local fermion. Just as in the bosonic case, we arrive at the fermionic $\mathbb{Z}_q$ TO by condensing $f_{\ell}^\dagger f_{\ell+1}$ for all $\ell$  with equal magnitude. Thus, the physical defects that one must proliferate to transition from the WMI to the $\mathbb{Z}_q$ TO are identical for the bosonic (see e.g.\ Fig.\ \ref{fig:bosonic_third_vortices}) and fermionic case for odd $p$ and $q$.

The field theory for the transition can be described as follows. We define the bosonic defect $b_{\ell} = f_{\ell}^\dagger c^\dagger$, i.e., $b^\dagger_{\ell}$ is the junction $f_{\ell}$ with the junction site filled with the local fermion. 
Thus, we could equivalently condense $b_{\ell}^\dagger b_{\ell+1}$ to arrive at the $\mathbb{Z}_q$ TO. As explained in Ref.\ \cite{musser_et_al_2025}, we can think of the fermionic WMI as a trivial fermionic band insulator at filling $1$ and a bosonic insulator of Cooper pairs of holes at filling $(q-p)/(2q)$. Then, the transition to the $\mathbb{Z}_q$ TO is identical to the discussion above Eq.\ \eqref{eq:zqto}. The bosonic defect $b_{\ell}$ becomes the $e$ particle with charge $p/q-1$.

\subsubsection*{Even \texorpdfstring{$p$}{p}}
Now, because $p$ is even, the product of the junctions yields the bosonic operator $(c^\dagger)^p$, and because $q$ is odd, each junction must be bosonic. Thus, we directly label each junction by bosonic fields $b_{\ell}$, where again $\ell$ ranges from $0$ to $q-1$. We now have the relation $(c^\dagger)^{p} = \prod_{\ell} b_{\ell}$, and we arrive at the fermionic $\mathbb{Z}_q$ TO by condensing $b_{\ell}^\dagger b_{\ell+1}$ for all $\ell$ with equal magnitude. Again, the physical defects that one must proliferate to transition from the WMI to the $\mathbb{Z}_q$ TO are identical for the bosonic (see e.g.\ Fig.\ \ref{fig:bosonic_third_vortices}) and fermionic case for even $p$ and odd $q$. Moreover, the field theory of this transition is equivalent to the transition from a bosonic insulator of Cooper pairs of electrons at filling $(p/2)/q$ to the minimal TO. Then, the transition to the $\mathbb{Z}_q$ TO is identical to the discussion above Eq.\ \eqref{eq:zqto}.

We have thus described the quantum melting of the WMIs into the minimal TO in terms of proliferation of topological defects of the WMIs. The elementary defects $b_{\ell}$ in both the even and odd $p$ cases serve as precursors to the $e$ particles in the QCL phase. 

\subsection{Even \texorpdfstring{$q$}{q}} \label{sec:even_q_fermion}
We now turn to the case of even $q$ (and because $p$ is coprime with $q$, odd $p$), where the minimal QCL exhibits $\mathbb{Z}_{2q}$ TO. Again, without loss of generality, we consider a period $q$  ordered 
WMI. If we follow the argument of Sec.\ \ref{sec:odd_q}, we have an even number of elementary junctions whose product must yield $p$ copies of the local anti-fermion. Since $p$ is odd, these junctions cannot be bosons even after attaching to them the local fermion. Thus, they cannot serve as the precursors to the $e$ particles in the QCL and thus the proliferation of composites of these defects cannot realize the minimal TO. We suspect that while these defects can be created, there is some obstruction to their proliferation.

This obstruction raises another natural question: what is the simplest WMI state that can quantum melt into the minimal TO? Motivated by our results for the odd $q$ case, we consider the $2q$-fold WMI. 
We later show in Section~\ref{sec:evenqQCL} that a $2q$-fold WMI arises naturally from transitions out of the QCL, and that no minimal $q$-fold WMI can be accessed by approaching the transition from the QCL side.

For a $2q$-fold WMI with filling $p/q$ and unit cell size $r'\times s'$ where $r's'= 2q$, we now have $2q$ defects $d_{\ell}$ (where now $\ell$ ranges from $0$ to $2q-1$ respectively). These defects can either be bosonic or fermionic. The product of all these defects yields $2p$ copies of the local antifermion $(c^{\dagger})^{2p} =  \prod_{\ell=0}^{2q-1} d_{\ell}$, in contrast to the odd-$q$ case. As before, condensing $d_\ell^\dagger d_{\ell+1}$ for all $\ell$ with equal magnitude yields the minimal QCL with $\mathbb{Z}_{2q}$ TO. 
The quantum melting scenario here is analogous to that of bosons at filling $\frac{p}{2q}$. The analogy is made more explicit upon forming electron pairs by smoothly bringing them together in each domain of the defects. These defects then correspond to those found in the bosonic Cooper pair case.
We note however, that our framework does not rely on any assumption of Cooper pairing of electrons.

\section{View from the fermionic QCL}\label{sec:QCL}
Having described the quantum melting transitions from the minimal fermionic WMI to the QCL in terms of the defects of the WMI, we now provide a complementary description of the transitions approaching from the minimal TO. Our description is in terms of the excitations of the QCLs. We again consider odd and even $q$ separately.

\subsection{Odd \texorpdfstring{$q$}{q}}
The QCL obtained by melting the WMI in this case is a symmetry enriched topological (SET) phase exhibiting $\mathbb{Z}_q$ topological order enriched by the global $U(1)$ charge conservation symmetry. The excitations are composed out of $q^2$ anyons of the $\mathbb{Z}_q$ TO and a local fermion, which is the electron. 
The $e$ particle carries physical charge $p/q$ and the ground state can be described as having one $e$ particle on each site, respecting both the translational symmetries and the filling constraint. 
The WMI, on the other hand, is a conventional phase of matter with broken lattice translation symmetries and no deconfined excitations.
To drive the QCL to a WMI, we must confine all the anyons while keeping the charge degree of freedom gapped. 
We achieve this by condensing a particle that is charge neutral and braids non-trivially with the other anyons. 
The charge-neutrality condition ensures that the charge remains gapped at the transition.

The $m$-particle of the $\mathbb{Z}_q$ TO is charge neutral and acquires a phase $2\pi/q$ when braided around an $e$ particle, and more generally a phase $2\pi j/q$ when braided around $e^j$.
The condensation of $m$ confines all the $e^j$ excitations and all their composites $e^j m^k$ since $m^k$ becomes identifed with the vacuum. 
Thus, the condensation of $m$ leads to an insulating state with all the anyons confined. 
LSMOH constraints then demand that the state break translational symmetries and be a WMI.

We now address the lattice periodicity of the $m$-condensed phase and whether it matches that of the WMIs discussed in the previous sections. Since the QCL consists of $e$-particles localized on sites, the $m$-particle sees a $2\pi/q$ flux per unit cell, and therefore, transform projectively under translations: $T_xT_y = \omega_q T_y T_x$, where $\omega_q \equiv \text{exp}(2\pi i /q)$. 
The magnetic unit cell for $m$ is thus enlarged $q$-times, introducing $q$ species of $m$-particles. Let us represent them by fields $\hat{\vphi}_{l}$. Following Ref.~\cite{balents_et_al_2005}, the projective action of translations on the $m$-particles can be implemented in the following way.
\begin{align}
    T_x&: \hat{\vphi}_{l} \rightarrow \hat{\vphi}_{l+1} \nonumber \\
    T_y&: \hat{\vphi}_l \rightarrow \hat{\vphi}_l \omega_q^{-l}
    \label{eq:projective}
\end{align}
To diagnose the translational symmetry breaking orders in the $m$-condensed phase, one can define the following gauge-invariant density-wave order parameters at wave-vector $\vec Q = \frac{2\pi}{q} (u,v)$, where $u$ and $v$ are integers.
\begin{align}
    \rho_{mn} = \sum_{l=0}^{n-1} (\hat{\vphi}_l)^\star \hat{\vphi}_{l+v} \omega_q^{lu}
\end{align}
The $m$-condensed phase with $\hat{\vphi}$ condensing into a generic complex vector results in density-wave order with an enlarged unit cell.
Stripe phases, phases with unit cells of the form $1\times q$ or $q\times 1$, can be described by the condensate: $\< \hat{\vphi}_l \> = \delta_{l,l_0} v_0$ (vertical stripes) and $\< \hat{\vphi}_l \> = \omega_q^{k_0l} v_0$ (horizontal stripes), where $k_0$ and $l_0$ are integers that label the $\mathbb{Z}_q$ clock order parameter and $v_0$ is a nonzero complex number.
For composite $q = rs$, where $r$ and $s$ are integers, we can also obtain WMI phases with wave-vectors $\frac{2\pi}{q} (r,0)$ and $\frac{2\pi}{q} (0,s)$ with the condensate
\begin{equation}
    \< \hat{\vphi}_l \> = \delta^r_{\ell,n} \omega_q^{m(l-n)} v_0,
\end{equation}
where $\delta^r_{\ell,n}$ is the Kronecker delta where $\ell-n$ is evaluated modulo $r$, and $(n,m)$ label the $\mathbb{Z}_r\times \mathbb{Z}_s$ order parameter. 
Thus, a transition out of the minimal TO to any of the possible minimal fermionic WMIs is allowed via $m$-condensation.

\subsection{Even \texorpdfstring{$q$}{q}}\label{sec:evenqQCL}
We now turn to the case where $q$ is even.  For even $q$, the minimal allowed QCL has $\mathbb{Z}_{2q}$ TO. The ground state is described by $e$ particles, which have physical charge $p/q$, localized on the sites. Ref.~\cite{musser_et_al_2025} details an interpretation in terms of localized fractionalized Cooper pairs. 
The $m$-particles now see a flux $\pi /q$ per unit cell, leading to $2q$ distinct species.
As in the odd-$q$ case, a confinement transition can be driven by condensing $m$. 
However, due to the projective action of translations on the $m$-particles, the resulting confined phase exhibits CDW order with a unit cell enlarged by at least a factor of $2q$.

We emphasize here that condensing $m$ is the only mechanism to fully confine \textit{all} the anyons of the TO while keeping a finite charge gap. 
Condensing $m^2$ does result in density-wave order with a unit cell that is enlarged by a factor of only $q$ because $m^2$ experiences a background flux of $2\pi/q$ per unit cell and obeys the transformation properties under translations specified in Eq.~\ref{eq:projective}. 
Crucially, however, the resulting phase is not a conventional WMI. Rather, it has residual $\mathbb{Z}_2$ topological order with the deconfined excitations: $\{1,e^q,m,e^q m \}$, which all braid trivially with $m^2$.
Condensation of other powers of $m$ that are not coprime with $2q$ similarly yields phases with both topological order and broken translational symmetry.
Importantly, while minimal WMI states are allowed at this filling, there can be no continuous phase transition from the QCL with $\mathbb{Z}_{2q}$ TO to these WMI states.

\section{Discussion}\label{sec:discussion}
In this paper, we consider defect-mediated continuous transitions at filling $p/q$, from fermionic WMIs to QCLs with the minimal TO. We show that while such a transition is possible from the minimal, i.e., $q$-fold, fermionic WMI, for $q$ odd, it is not possible for $q$ even. This result has implications for a variety of experimental and theoretical directions. 

On the experimental front, performing STM imaging, similarly as Ref.\ \cite{Xiang2025_quantum_melting_Wigner} of the transition between a fermionic WMI and metallic liquid may reveal intermediate gapped insulating states without charge order. Our work constrains the topological orders that state may have. In particular, at filling $1/2$, the minimal fermionic WMI cannot directly transition into a topologically ordered state. For even the minimal topological order ($\mathbb{Z}_4$) to be realized, the minimal WMI must first transition into at least a $4$-fold WMI. Because the minimal WMI is the favored charge ordering pattern in many physical systems with strong Coulomb repulsion, our results suggest that odd-$q$ fillings are favorable for the search of QCL states. Such constraints also appear in the theoretical search for lattice models of fermionic QCLs and transitions into those QCLs. In particular, extending the work of Ref.\ \cite{zhou_et_al_2024}, which considered spin-full electrons at $1/3$ filling on a 2d lattice, to spinless fermions at $1/3$ filling may reveal could prove fruitful in the search for fermionic gapped QCLs. 

Certainly many questions about 2d electronic transitions persist as well. For example, transitions between the fermionic QCL and the Fermi-liquid metal are not well understood. Additionally, this paper's defect proliferation approach could be used to study the melting of electronic crystals that spontaneously break continuous translation symmetry. This question has attained recent importance due to the discovery of integer and fractional anomalous Hall effects in rhombohedral graphene \cite{Lu2024,Lu2025}. Explaining the rich phase diagram of 2d materials like rhombohedral graphene through proliferation of crystalline defects with nontrivial quantum numbers (see e.g. Ref.\ \cite{Cho_et_al_2015,manjunath_barkeshli_2021}) is an open question. 

Finally, as discussed in Sec.\ \ref{sec:bosonic_q_greater_2}, the transition from the WMI to the QCL can be thought of as the proliferation of particular sets of domain walls. While a generalized symmetry perspective has been used to describe transitions out of states that spontaneously break an internal symmetry \cite{Pace_2024_Emergent_GS} through the proliferation of domain walls, it remains to be seen how LSMOH constraints affect this story. 

\section*{Acknowledgements}
We are grateful to Aprem Joy, Max Metlitski, Sal Pace, and Zhengyan Darius Shi for discussions. We thank Seth Musser for feedback on a previous version of the manuscript. 
The work of A. Krishnan was supported
by the National Science Foundation Graduate Research Fellowship under Grant No. 1745302. A. Krishnan
also acknowledges support from the Paul and Daisy Soros Fellowship and the Barry M. Goldwater
Scholarship Foundation. 
A. Kumar was supported by the Gordon and Betty Moore Foundation EPiQS Initiative through Grant No. GBMF8684 at the Massachusetts Institute of Technology and by the Deutsche Forschungsgemeinschaft (DFG) through CRC1238 (Project No. 277146847, project C02).
T.S. is  supported by NSF grant DMR-2206305.

\appendix
\section{Choice of basis for the vortex fields} \label{app:vortex_basis}
In this appendix, we first argue that some choices of basis for the superfluid vortex fields are more convenient than others when considering the dual bosonic parton theory. We then show how to construct those bases for different unit cell patterns. 

\subsection{Dependence of parton description on the choice of basis}\label{app:explanation}
For ease of explanation, let us consider filling $1/3$ and show that the superfluid vortex basis in Ref.\ \cite{balents_et_al_2005} is not convenient for describing the WMI to superfluid transition as the proliferation of $q$ partons. In Ref.\ \cite{balents_et_al_2005}, the description of the superfluid-Mott transition in terms of superfluid vortices is given by the following Lagrangian:
\begin{equation}
    \mathcal{L} = \mathcal{L}[\tilde{\varphi}_0,b] +  \mathcal{L}[\tilde{\varphi}_1,b]+  \mathcal{L}[\tilde{\varphi}_2,b] +V_v(\vec{\tilde{\varphi}}) + \frac{1}{2\pi} b \wedge d A,
\end{equation}
where 
\begin{equation}
    T_x: \tilde{\varphi}_\ell \to \tilde{\varphi}_{\ell+1}, \quad T_y: \tilde{\varphi}_\ell \to \omega^{-\ell} \tilde{\varphi}_\ell,
\end{equation}
 $\omega = e^{2\pi i /3}$, $b$ is an internal $U(1)$ gauge field, and $V_v$ is a vortex potential that respects the $T_x$ and $T_y$ symmetries. The superfluid phase is described by all three $\tilde{\varphi}$ fields being gapped. To break the $T_x$ symmetry, we drive the system through a transition where one of the $\tilde{\varphi_\ell}$ fields spontaneously condenses.
The order parameter that detects breaking of $T_x$ is 
\begin{equation}
    \mathcal{O}_x = |\tilde{\varphi}_0|^2 + \omega |\tilde{\varphi}_1|^2 + \omega^2 |\tilde{\varphi}_2|^2. 
\end{equation}
The particle-vortex dual description of this theory is 
\begin{equation}
    \mathcal{L} = \mathcal{L}[\tilde{\phi}_0,a_0] +  \mathcal{L}[\tilde{\phi}_1,a_1]+  \mathcal{L}[\tilde{\phi}_2,a_2] +V(\vec{\tilde{\phi}}) + \frac{1}{2\pi} b \wedge d (A+a_0 +a_1 + a_2),
\end{equation}
where now
\begin{equation}
    T_x: \tilde{\phi}_\ell \to \tilde{\phi}_{\ell+1}, \quad T_y: \tilde{\phi}_\ell \to  \tilde{\phi}_\ell,
\end{equation}
$a_\ell$ is an internal $U(1)$ gauge field,
$T_x$ and $T_y$ has the same action on $a_\ell$, and $V$ is a potential that respects the lattice translation symmetries. When a single $\tilde{\varphi}_\ell$ field is condensed, the corresponding $\tilde{\phi}_\ell$ field is gapped and the other two $\tilde{\phi}_{\ell'}$ fields are condensed.
Now, the order parameter that detects breaking of $T_x$ is 
\begin{equation}
    \mathcal{O}_x = -|\tilde{\phi}_0|^2 - \omega |\tilde{\phi}_1|^2 - \omega^2|\tilde{\phi}_2|^2.
\end{equation}
Because $1+\omega + \omega^2 = 0$, the two $\mathcal{O}_x$ order parameters are equivalent. Moreover, in the superfluid phase, $\mathcal{O}_x$ is zero, so the magnitudes of all $\tilde{\phi}_\ell$ fields must be equivalent.

In the superfluid vortex description, the WMI to superfluid transition is described by gapping the previously condensed $\tilde{\varphi}_\ell$ field through the condensation of $\mathcal{M}_b$, the monopole of the internal gauge field $b$. In the bosonic parton description, the WMI to superfluid transition is described by two simultaneous processes: (1) the condensation of the previously gapped $\tilde{\phi}_\ell$ field and (2) the locking of $|\tilde{\phi}_\ell|$ to $|\tilde{\phi}_{\ell'}|$ in order to set $\mathcal{O}_x$ to zero. Because the WMI to superfluid transition is described by these two simultaneous processes, a description in terms of the proliferation of partons (or defects of the WMI) is not natural in this basis. 

However, if instead we choose a basis where the $T_x$ broken phase is described by the condensation of all $\varphi_\ell$ fields locked to each other, as done in Sec.\ \ref{sec:vertical_stripe_WMIs}, then, in the dual description, all $\phi_\ell$ fields are gapped and the WMI to superfluid transition can be described by the proliferation of all $\phi_\ell$ fields. 

\subsection{Choice of basis for various unit cells shapes} \label{app:basis_choice}
We first discuss $q\times 1$ unit cells. In the main text, we choose a basis for $\varphi_\ell$ where the action of $T_x$ and $T_y$ are given by Eq.\ \eqref{eq:q_odd_psg} for q odd and Eq.\ \eqref{eq:q_even_psg} for q even. We explicitly construct this basis from the $\tilde{\varphi}_\ell$ basis in Ref. \cite{balents_et_al_2005}.

For $q$ odd, we define 
\begin{equation}
    \varphi_\ell = \frac{1}{\sqrt{q}}\sum_{n=0}^{q-1} \omega^{-(\ell-n)(\ell - n+1)/2}\tilde{\varphi}_n. 
\end{equation}
For $q$ even, we define 
\begin{equation}
    \varphi_\ell = \frac{1}{\sqrt{q}}\sum_{n=0}^{q-1} \omega^{-(\ell-n)^2/2}\tilde{\varphi}_n. 
\end{equation}

Now, we discuss $r\times s$ unit cells. It is impossible to choose a basis where both $T_x$ and $T_y$ act as clock or shift operators and condensing a vector of $\varphi_\ell$ with all nonzero entries results in a symmetry breaking pattern with $r\times s$ unit cells. However, we can choose a basis $\vphi_\ell'$ where $T_x$ still acts as a shift operator, $T_y$ has a more complicated action on the $\varphi_\ell'$ fields, and condensing a vector of $\varphi_\ell'$ with all nonzero entries results in a symmetry breaking pattern with $r\times s$ unit cells. First, we note that in the $\tilde{\varphi}_\ell$ basis in Ref.\ \cite{balents_et_al_2005}, the condensate that results in an $r\times s$ unit cell is 
\begin{equation}
  \ev{\tilde{\varphi}_\ell} = \delta^r_{\ell,n} \omega^{m(\ell - n)} v_0,
\end{equation}
where $v_0$ is some nonzero complex number, $\delta^r_{\ell,n}$ is the Kronecker delta where $\ell-n$ is evaluated modulo $r$, and $(n,m)$ label the $\mathbb{Z}_r \times \mathbb{Z}_s$ order parameter. We now need to find a unitary transformation $U$ of $\tilde{\varphi}$ basis that preserves $T_x$, and satisfies the property that condensing a vector of $\varphi_\ell'$ with all nonzero entries results in a symmetry breaking pattern with $r\times s$ unit cells. The first property is equivalent to requiring that $U$ commute with $T_x$ and thus be a circulant matrix, i.e., 
\begin{equation}
    U = F_q^\dagger \text{diag}(e^{i\theta_0}, \ldots, e^{i\theta_{q-1}}) F_q, \quad F_{q,jk} = \frac{1}{\sqrt{q}} \omega^{jk},
\end{equation}
where $\theta_n$ can be any angle.
In other words, 
\begin{equation}
    U_{jk} = \frac{1}{q} \sum_{n} e^{i\theta_n} \omega^{n(k-j)},
\end{equation}
where $k-j$ is evaluated modulo $q$.
The second property is the requirement that for all condensation patterns $U\ev{\vec{\tilde{\varphi}}}$ is nonzero. The $j$th entry of this vector is 
\begin{align}
   \begin{split}
   \ev{\varphi_j'} &= U_{j\ell}\tilde{\varphi}_\ell\\
   &=\frac{1}{q} \sum_{k=0}^{q-1} \sum_{\ell=0}^{q-1} e^{i\theta_k} \omega^{k(\ell-j)} \delta^r_{\ell,n} \omega^{m(\ell - n)} v_0\\
   &= \frac{1}{q} \sum_{k=0}^{q-1} \sum_{t=0}^{s-1} e^{i\theta_k} \omega^{k(n + tr-j)} \omega^{m tr} v_0
   \end{split}
\end{align}
Choosing different irrational $\theta_k$ should thus ensure that each $\ev{\varphi_j'}$ is nonzero. The action of $T_y$ on this basis is generally complicated, as $UT_yU^{-1}$ is not guaranteed to be a clock or shift operator.

\bibliography{bibliography-short.bib}
\bibliographystyle{JHEP}

\end{document}

%% file: defs.tex
\newcommand{\Z}{\mathbb{Z}}

\newcommand{\beq}{\begin{equation}}
\newcommand{\eeq}{\end{equation}}
\newcommand{\ba}{\begin{array}{ccc}}
\newcommand{\ea}{\end{array}}
\newcommand{\bea}{\begin{eqnarray}}
\newcommand{\eea}{\end{eqnarray}}

\usepackage{simpler-wick}